\documentclass[10pt]{article}
\usepackage{graphicx}
\usepackage{textcomp}
\usepackage{amssymb}
\usepackage{amsmath}
\usepackage{epstopdf}
\usepackage{epsfig}
\usepackage{multicol}
\begin{document}

\begin{center}
\textbf{Simply Explicitly Invertible Approximations to 4 Decimals of Error Function and Normal Cumulative Distribution Function}
\end{center}

\noindent
\begin{center}
\textbf{Alessandro Soranzo}\\
Dipartimento di Matematica e Informatica -- Universit\`a degli Studi di Trieste\\
Trieste -- Italy -- e-mail:soranzo@units.it\\
\end{center}

\begin{center}
\textbf{Emanuela Epure}\\
Esteco S.R.L -- Area Science Park -- Trieste -- Italy -- e-mail: epure@esteco.com\\

\end{center}

\noindent
\textbf{Abstract.} We improve the Winitzki's Approximation of the error function 
$erf(x)\cong \sqrt{1-e^{-x^2\frac{\frac{4}{\pi}+0.147x^2}{1+0.147x^2}}}$ which has error $|\varepsilon (x)| < 1.25 \cdot 10^{-4}$ $\forall x \ge 0$ till reaching 4 decimals of precision with $|\varepsilon (x)| < 2.27 \cdot 10^{-5}$; also reducing slightly the relative error. Old formula and ours are both explicitly invertible, essentially solving a biquadratic equation, after obvious substitutions. 
Then we derive approximations to 4 decimals of normal cumulative distribution function $\Phi (x)$, of erfc$(x)$ and of the $Q$ function (or cPhi). 



\noindent
In this note we improve the Winitzki's Approximation of the error function till reaching 4 decimals of precision both for erf and $\Phi$, reducing about 5.5 times the respective absolute errors (and reducing the relative errors too).

\noindent
\textbf{Lemma (Winitzki's Approximation of erf).} (See \cite{Winitzki})

$$
erf(x)\cong \sqrt{1-e^{-x^2\frac{\frac{4}{\pi}+0.147x^2}{1+0.147x^2}}}
\quad |\varepsilon (x) |<1.25\cdot 10^{-4} 
\quad |\varepsilon_r (x) |<1.28\cdot 10^{-4} 
\,\, \forall x \ge 0
$$

\noindent
\textbf{Theorem (Improving of Winitzki's Approximation of erf).}
\begin{equation}
\label{ourErf}
erf(x) \cong \sqrt{1-{e}^{\frac{-1.2735457x^2- 0.1487936x^4}{1+0.1480931x^2+0.0005160x^4}}}
\end{equation}

$$|\varepsilon (x)| <2.27\cdot 10^{-5} \,\, \forall x \ge 0 
\qquad |\varepsilon_r (x)| <1.21\cdot 10^{-4} \,\, \forall x \ge 0$$

\noindent
\textit{Proof (only for absolute error).}
For $0\le x\le 4$, see the figure above
$^[$\footnote{The figure represents the graph of
$2.27\cdot 10^{-5}- \Big| erf(x)-\sqrt{1-{e}^{\frac{-1.2735457x^2- 0.1487936x^4}{1+0.1480931x^2+0.0005160x^4}}} \Big|$, made by the software Mathematica, showing that the quantity is positive.

\noindent
To verify that the graph do not intersect the $x$ axis, you may make zooms in the subdomains $[0.2, 0.3]$, $[0.75, 0.85]$, $[1.35, 1.45]$ and $[2.1, 2.2]$.}$^]$.

\noindent
For $x>4$ let's consider that
it is $erf(4)= 0.99999998458...$ and $erf (x)\rightarrow 1$ and erf is increasing, then

\begin{equation}
\label{eq:unoMenoErf}
\big( \forall x>4 \big)\qquad |1-erf(x)|<10^{-7} .
\end{equation}

\noindent
Let's $\eta (x)$ our approximation of erf$(x)$ as in (\ref{ourErf}):
$erf(x)\cong \eta (x):= \sqrt{1-e^{E(x)}}$
being $E(x):=\frac{-1.2735457x^2- 0.1487936x^4}{1+0.1480931x^2+0.0005160x^4}$.
It is:

$$\big( \forall x>4 \big) \quad 0<3=\Big(\frac{1}{3}15-1\Big)^2-1-12<
(\frac{1}{3}\cdot 4^2-0.795)^2-0.795^2-12<$$

$$=\Big(\frac{1}{3}x^2-\frac{3}{2}0.53 \Big)^2-\Big( \frac{3}{2}0.53 \Big)^2-12=
\frac{1}{9} x^4-0.53x^2-12<$$
$$<0.14176x^4-0.53x^2-12=(0.148x^4-0.00624)x^4+(1.27x^2-1.8)x^2-12$$

$$\Rightarrow 1.27x^2+0.148x^4>12+1.8x^2+0.00624x^4$$
$$\Rightarrow \frac{1.27x^2+0.148x^4}{12}>1+0.15x^2+0.00052x^4$$

$$\Rightarrow E(x)=\frac{-1.2735457x^2- 0.1487936x^4}{1+0.1480931x^2+0.0005160x^4}<
\frac{-1.27x^2-0.148x^4}{1+0.15x^2+0.00052x^4}<
-12$$

$$\Rightarrow e^{R(x)} <e^{-12}<e^{-12}+(e^{-12}-e^{-24})=2e^{-12}-e^{-24}=1-(1-e^{-12})^2$$

$$\Rightarrow (1-e^{-12})^2<1-e^{R(x)} \quad \Rightarrow \quad 1-e^{-12}<\sqrt{1-e^{R(x)}}$$

$$\Rightarrow 0<1-\eta (x)=1- \sqrt{1-e^{R(x)}} <e^{-12}\Rightarrow$$

\begin{equation}
\label{eq:unoMenoEta}
\big( \forall x>4 \big)\qquad |1-\eta (x)|<e^{-12}.
\end{equation}

\noindent
By (\ref{eq:unoMenoErf}) and (\ref{eq:unoMenoEta}) it is

$$\big( \forall x>4 \big)\qquad |erf(x)-\eta (x)|\le |1-erf (x)|+|1-\eta (x)| < 10^{-7}+e^{-12}<10^{-5}.$$

\noindent
$\square$

\begin{center}
\begin{figure}
\includegraphics[height=3.2 cm,width=6.4 cm]{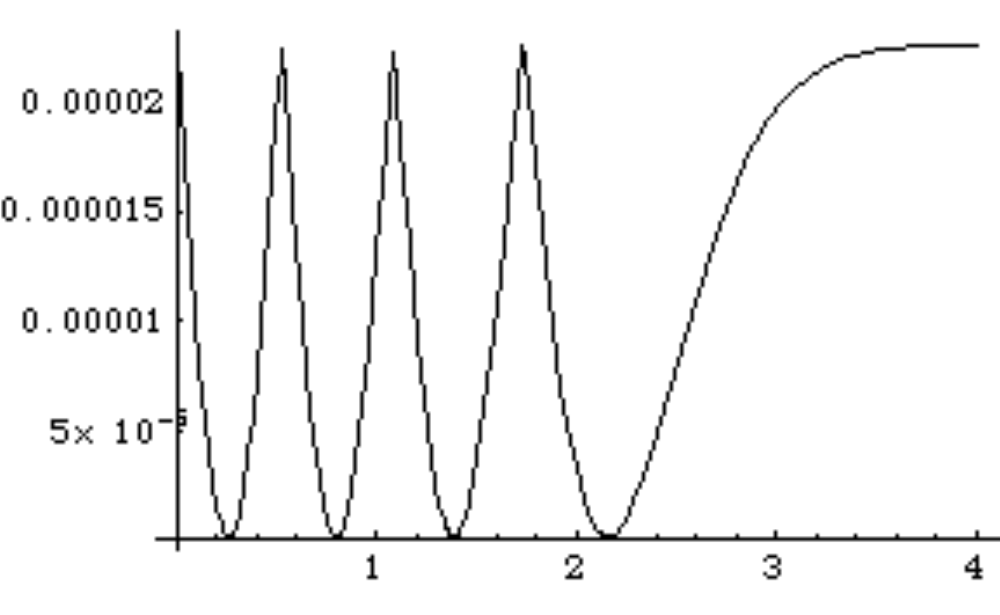}
\label{fig:WinitzkiProof.pdf}
\end{figure}
\end{center}

\newpage
\begin{center}
\begin{Large}
\textbf{TABLE ERF2.27E-5 }
\end{Large}

\end{center}

\noindent
\textbf{Simply explicitly invertible approximations to 4 decimals of the error function erf, of complementary error function
erfc, of the normal cumulative distribution function $\Phi$, and of Q function, for $x\ge 0$.}

\noindent
\textbf{With majorizations of absolute and relative errors, and definitions.}

$$
erf(x):= \int_{0}^{x} \frac{2}{\sqrt {\pi}} e^{-t^2} dt=1-erfc(x)
\quad \quad
\Phi (x):=\int_{-\infty}^x \frac{1} {\sqrt{2\pi}}e^{-\frac{t^2}{2}}=1-Q(x)
$$

\begin{small}

\begin{center}
$------------------------------------------$
\end{center}


\begin{multicols}{2}
\noindent
$A.$
$ \qquad erf(x) \cong$
$$ \cong \sqrt{1-{e}^{\frac{-1.2735457x^2-0.1487936x^4}{1+0.1480931x^2+0.0005160x^4}}}$$

\vfil
\columnbreak
\setlength{\columnseprule}{0.5pt}
\noindent

\begin{center}
$|\varepsilon (x)| <2.27\cdot 10^{-5} \,\, \forall x \ge 0$
$\: ^[$\footnote{For $x\gtrsim 4.125$ the approximation $erf (x)\cong 1$ has less absolute error.}$^]$
\end{center}

\begin{center}
$|\varepsilon_r (x)| <1.21\cdot 10^{-4} \,\, \forall x \ge 0$
\end{center}

\end{multicols}

\begin{center}
$------------------------------------------$
\end{center}

\begin{multicols}{2}
\noindent
$B.$
$\qquad erfc(x)\cong$
$$\cong 1-\sqrt{1-{e}^{\frac{-1.2735457x^2- 0.1487936x^4}{1+0.1480931x^2+0.0005160x^4}}}$$

\vfil
\columnbreak
\setlength{\columnseprule}{0.5pt}
\noindent
\begin{center}
$|\varepsilon (x)| <2.27\cdot 10^{-5} \,\, \forall x \ge 0$
$\:^[$\footnote{For $x\gtrsim 4.125$ the approximation $erfc (x)\cong 0$ has less absolute error.}$^]$
\end{center}

\begin{center}
$|\varepsilon_r (x)| <1\% \,\, \forall x \in [0,b],\,b>2.1588$
\end{center}

\end{multicols}

\begin{center}
$------------------------------------------$
\end{center}

\begin{multicols}{2}
\noindent
$C.$
$\qquad \Phi(x) \cong$
$$\cong \frac{1}{2} + \frac{1}{2} \sqrt{1-{e}^{\frac{-1.2735457x^2-0.0743968x^4}{2+0.1480931x^2+0.0002580x^4}}}$$

\vfil
\columnbreak
\setlength{\columnseprule}{0.5pt}
\noindent

\begin{center}
$|\varepsilon (x)| <1.14\cdot 10^{-5} \,\, \forall x \ge 0$
$\: ^[$\footnote{For $x\gtrsim 5.834$ the approximation $\Phi (x)\cong 1$ has less absolute error.}$^]$
\end{center}

\begin{center}
$|\varepsilon_r (x)| <1.78\cdot 10^{-5} \,\, \forall x \ge 0$
\end{center}

\end{multicols}

\begin{center}
$------------------------------------------$
\end{center}

\begin{multicols}{2}
\noindent
$D.$
$\qquad Q(x)\cong$
$$ \cong \frac{1}{2} - \frac{1}{2} \sqrt{1-{e}^{\frac{-1.2735457x^2-0.0743968x^4}{2+0.1480931x^2+0.0002580x^4}}}$$

\vfil
\columnbreak
\setlength{\columnseprule}{0.5pt}
\noindent

\begin{center}
$|\varepsilon (x)| <1.14\cdot 10^{-5} \,\, \forall x \ge 0$
$^[$\footnote{For $x\gtrsim 5.834$ the approximation $Q (x)\cong 0$ has less absolute error.}$^]$
\end{center}

\begin{center}
$|\varepsilon_r (x)| <1\% \,\, \forall x \in [0,b],\,b>3.053$
\end{center}
\end{multicols}
\end{small}


\end{document}